\documentclass[prl,twocolumn]{revtex4}
\usepackage{graphicx}

\begin{document}

\title{Coherence Phenomena in the Phase-sensitive Optical
Parametric Amplification inside a Cavity}
\author{Hongliang Ma, Chenguang Ye, Dong Wei, Jing Zhang$^{\dagger }$}
\affiliation{The State Key Laboratory of Quantum Optics and Quantum
Optical Devices, Institute of Opto-Electronics, Shanxi University,
Taiyuan 030006, P.R. China}

\begin{abstract}
We theoretically and experimentally demonstrate coherence phenomena
in optical parametric amplification inside a cavity. The mode
splitting in transmission spectra of phase-sensitive optical
parametric amplifier is observed. Especially, we show a very narrow
dip and peak, which are the shape of $\delta $ function, appear in
the transmission profile. The origin of the coherence phenomenon in
this system is the interference between the harmonic pump field and
the subharmonic seed field in cooperation with dissipation of the
cavity.
\end{abstract}

\maketitle

$Introduction$\textit{.} --- Coherence and interference effects play
very important roles in determining the optical properties of
quantum systems. Electromagnetically induced transparency (EIT)
\cite{one} in quantum-mechanical atomic systems is a well understood
and thoroughly studied subject. EIT has been utilized in a variety
of applications, such as lasing without inversion \cite{two}, slow
and stored light \cite {two1,two2}, enhanced nonlinear optics
\cite{three}, and quantum computation and communication \cite{four}.
Relying on destructive quantum interference, EIT is a phenomenon
where the absorption of a probe laser field resonant with an atomic
transition is reduced or even eliminated by the application of a
strongly driving laser to an adjacent transition. Since EIT results
from destructive quantum interference, it has been recently
recognized that similar coherence and interference effects also
occur in classical systems, such as plasma \cite{five}, coupled
optical rsonators\cite {six}, mechanical or electric oscillators
\cite{seven}. In particular, the phenomenology of the EIT and
dynamic Stark effect is studied theoretically in a dissipative
system composed by two coupled oscillators under linear and
parametric amplification using quantum optics model in Ref.
\cite{eight}. The classical analog of EIT is not only helpful to
understand deeply the physical meaning of EIT phenomenon, but also
offers a number of itself important
applications, such as slow and stored light by coupled optical resonators%
\cite{nine}.

In this Letter, we extend the model in Ref. \cite{eight} and present
a new system - phase-sensitive optical parametric amplifier (OPA) to
demonstrate coherence effects theoretically and experimentally. We
observe mode splitting in transmission spectra of OPA. Especially,
we show a very narrow dip and peak, which are the shape of $\delta $
function, appear in the transmission profile. This phenomenon
results from the interference between the harmonic pump field and
the subharmonic seed field in OPA. The destructive and constructive
interference correspond to optical parametric deamplifier and
amplifier respectively, which are in cooperation with dissipation of
the cavity. The absorptive and dispersive response of an optical
cavity for the probe field is changed by optical parametric
interaction in the cavity. Phase-sensitive optical parametric
amplifier presents a number of new characteristics of coherence
effects.

$Theoretical$ $model.$ --- Consider the interaction of two optical
fields of frequencies $\omega $ and $2\omega $, denoted by
subharmonic and harmonic wave (the pump), which are coupled by a
second-order, type-I nonlinear crystal in a optical cavity as shown
in Fig.1. The cavity is assumed to be a standing wave cavity, and
only resonant for the subharmonic field with dual-port of
transmission $T_{HR}$ and $T_c$, internal losses $A$ and length $L$
(roundtrip time $\tau =2L/c$). We
consider both the subharmonic seed beam $a^{in}$ and harmonic pump beam $%
\beta ^{in}$ are injected into the back port ($T_{HR}$ mirror) of
the cavity, where the relative phase between the injected field is
adjusted by a movable mirror outside the cavity. $T_{HR}$ mirror is
a high reflectivity mirror at the subharmonic wavelength, yet has a
high transmission coefficient at the harmonic wavelength and $T_c$
mirror has a high reflectivity coefficient for the harmonic wave.
The harmonic wave makes a double pass through the nonlinear medium.
The equation of motion for the mean value of the subharmonic
intra-cavity field can then be derived by the semiclassical method
\cite{ten} as
\begin{equation}
\tau \frac{da}{dt}=-i\tau \Delta a-\gamma a+g\beta ^{in}a^{\ast
}+\sqrt{2\gamma _{in}}a^{in}.  \label{degen}
\end{equation}
The decay rate for internal losses is $\gamma _l=A/2 $ and the
damping associated with coupling mirror and back mirror is $\gamma
_c=T_c/2 $ and $\gamma _{in}=T_{HR}/2 $, respectively. The total
damping is denoted by $\gamma =\gamma _{in}+\gamma _c+\gamma _l$.
$\Delta $ is the detuning between the cavity-resonance frequency
$\omega _c$ and the subharmonic field frequency $\omega $. The
strength of the interaction is characterized by the nonlinear
coupling parameter $g$. Eq.\ref{degen} is complemented
with the boundary conditions $a^{out}=\sqrt{2\gamma _c}a$ and $%
a^{ref}=-a^{in}+\sqrt{2\gamma _{in}}a$ to create propagating beams,
where $a^{out}$ is the transmitted field from the coupling mirror $T_c$ and $%
a^{ref}$ is the reflected field from the back mirror $T_{HR}$. The
phase-sensitive optical parametric amplifier always operates below
the threshold of optical parametric oscillation (OPO) $\beta
_{th}^{in}=\gamma /g$. Eq.\ref{degen} ignores the third-order
term\cite{eleven} describing the conversion losses due to harmonic
generation. For simplicity, we assume that the phase of the pump
field is zero in any case, i.e, $\beta ^{in}$ is real and positive
value. The intra-cavity field $a$
and the injected field $a^{in}$ are expressed as $a =$ $%
\alpha \exp \left( -i\phi \right) $ and $a^{in} =$ $%
A_{in}\exp \left( -i\varphi \right) $ respectively. Here, $\alpha $ and $%
A_{in}$ are real values, $\phi $ and $\varphi $ are the relative
phase between the intra-cavity field and the pump field as well as
between the seed field and the pump field, respectively. If the
harmonic pump is turned off, the throughput for the non-impedance
matched subharmonic seed beam is given by $ a_{no~pump}^{out}
=2\sqrt{\gamma _c\gamma _{in}}A_{in}/(\gamma +i\tau \Delta )$. The
subharmonic seed beam is subjected to either amplification or
de-amplification, depending on the chosen relative phase between the
subharmonic field and the pump field.

$Case1:$ Consider the transmitted intensity of the subharmonic seed beam as
a function of the detuning $\Delta $ between the subharmonic field frequency
and the cavity-resonance frequency, and keep the pump field of frequency $%
\omega _p=2\omega $ constant. Setting the derivative to zero
($d\alpha /dt=0$) and separating the real and image part of
Eq.\ref{degen}, the steady state solutions of the amplitude and
relative phase of the intra-cavity field are given by
\begin{eqnarray}
-\gamma \alpha +g\beta ^{in}\alpha \cos 2\phi +\sqrt{2\gamma _{in}}%
A_{in}\cos \left( \phi -\varphi \right) &=&0,  \label{steady} \\
-\tau\Delta \alpha +g\beta ^{in}\alpha \sin 2\phi +\sqrt{2\gamma _{in}}%
A_{in}\sin \left( \phi -\varphi \right) &=&0.  \nonumber
\end{eqnarray}
When the amplitude and relative phase of the subharmonic seed beam are
given, the transmitted intensity of the subharmonic beam is obtained from Eq.%
\ref{steady} and the boundary condition. Fig.2(a) shows a Lorentzian
profile of the subharmonic transmission when the pump field is
absent. This corresponds to the typical transmitted spectrum of the
optic al empty cavity. When the injected subharmonic field is out of
phase ($\varphi =\pi /2)$ with the pump field, the subharmonic
transmission profile is shown in Fig.2(b,c,d) for different pump
powers, in which there is a symmetric mode splitting. The
transmitted power of the subharmonic beam is normalized to the power
in the absence of the pump and zero detuning. The transmission
spectra show that the dip becomes deeper and two peaks higher as the
pump intensity increases. The origin of mode splitting in
transmission spectra of OPA is destructive interference in
cooperation with dissipation of the cavity. If the subharmonic field
resonates in the cavity perfectly, i.e. $\Delta =0$, the subharmonic
intra-cavity field and the pump filed are exactly out of phase and
will interfere destructively to produce the deamplification for the
subharmonic field in the nonlinear crystal. Thus a dip appears at
the zero detuning of the transmission profile. If the subharmonic
field is not quite resonant in the cavity perfectly, that is, the
subharmonic field's frequency is not exactly an integer multiple of
the free spectral range (but close enough to build up a standing
wave), the phase difference between the subharmonic intra-cavity
field and the pump field will not be exactly out of phase and will
increase as the detuning increases. The subharmonic intra-cavity
field will change from deamplification to amplification as the phase
difference increases. Thus we see that the transmission profile has
two symmetric peaks at two detuning frequencies. When the phase of
the injected subharmonic field is deviated from out of phase with
the pump field, i.e. $\varphi =\pi /2\pm \theta $, an asymmetric
mode splitting in the subharmonic transmission profile is
illustrated in Fig.2(e,f), in which the dip is deviated from the
zero detuning of the transmission profile and two peaks have
different amplitude.

$Case2:$ Consider the subharmonic transmission profiles when the
frequency of the pump field is fixed at $\omega _p=2(\omega
_c+\Omega )$. When scanning the frequency of the the subharmonic
seed beam, an idler field in the OPA cavity will be generated with
the frequency $\omega _i=\omega _p-\omega $ due to energy
conservation. The equation of motion of OPA become
frequency-nondegenerate and is given by

\begin{eqnarray}
\tau \frac{da}{dt} &=&-i\tau \Delta a-\gamma a+g\beta ^{in}a_i^{\ast }+\sqrt{%
2\gamma _{in}}a^{in},  \label{nondegen} \\
\tau \frac{da_i}{dt} &=&-i\tau \Delta _ia_i-\gamma a_i+g\beta
^{in}a^{\ast }  \nonumber
\end{eqnarray}
where $a_i$ is the idler field in the OPA cavity. $\Delta _i$ is the
detuning between the cavity-resonance frequency $\omega _c$ and the
idler field frequency $\omega _i$. Thus the subharmonic transmission
profile in this case is obtained from Eq.\ref {nondegen} for $\omega
\neq \omega _i$ and Eq.\ref{degen} for $\omega =\omega _i$. When
$\Omega =0$, so $\Delta =-\Delta _i$, the stationary solution of the
subharmonic and idle field is given by solving the mean-field
equations of Eq.\ref{nondegen} and using the input-output
formalisms. We obtain
\begin{eqnarray}
A^{out} &=&\frac{2\sqrt{\gamma _c\gamma _{in}}}{i\tau\Delta +\gamma -\frac{%
(g\beta ^{in})^2}{i\tau\Delta +\gamma }}A^{in}, \\
A_i^{out} &=&\frac{2\sqrt{\gamma _c\gamma _{in}}g\beta ^{in}}{\left(
-i\tau\Delta +\gamma \right) ^2-(g\beta ^{in})^2}A^{in*}.  \nonumber
\end{eqnarray}
We will record the total output power including the subharmonic and
idle field. The transmitted power of the subharmonic beam is given
by

\begin{equation}
P_{out}^{nor}=\left\{
\begin{array}{c}
\left| \frac \gamma {i\tau\Delta +\gamma -\frac{(g\beta
^{in})^2}{i\tau\Delta +\gamma }}\right| ^2+\left| \frac{\gamma
g\beta ^{in}}{\left( -i\tau\Delta
+\gamma \right) ^2-(g\beta ^{in})^2}\right| ^2\qquad \\
\qquad \qquad \qquad \qquad \mathrm{if}\quad \omega \neq \omega _i\quad \\
\frac{\gamma ^2}{(\gamma \pm g\beta ^{in})^2}\quad \qquad
\mathrm{if}\quad \omega =\omega _i.
\end{array}
\right.
\end{equation}
Here, $\pm $ corresponds to the deamplifier and amplifier in
frequency-degenerate OPA. Fig.3(a) and (b) show that the very narrow
dip and peak, which is the shape of $\delta $ function, appear in
the transmission profile. This novel coherence phenomena results in
that the destructive and constructive interference are established
only in the point of $\omega =\omega _i$, and completely destroyed
in the other frequencies.

$Experiment$. --- The experimental setup is shown schematically in
Fig.4. A
diode-pumped intracavity frequency-doubled continuous-wave(cw) ring Nd:YVO$_{%
\text{4}}$/KTP single-frequency green laser severs as the light
sources of the pump wave (the second-harmonic wave at $532$ $nm$)
and the seed wave (the fundamental wave at $1064$ $nm$) for OPA. The
green beam doubly passes the acousto-optic modulator (AOM) to shift
the frequency 440 MHz. The infrared beam doubly passes AOM to shift
the frequency around 220 MHz. We actively control the relative phase
between the subharmonic and the pump field by adjusting the phase of
the subharmonic beam with a mirror mounted upon a piezoelectric
transducer (PZT). Both beams are combined together by a dichroic
mirror and injected into the OPA cavity. OPA consists of
periodically poled KTiOPO$_4$ (PPKTP) crystal (12 $mm$ long) and two
external mirrors separated by $63$ $mm$. Both end faces of crystal
are polished and coated with an antireflector for both wavelengths.
The crystal is mounted in a copper block, whose temperature was
actively controlled at millidegrees
kelvin level around the temperature for optical parametric process (31.3${%
{}^{\circ }}$C). The input coupler M1 is a $30$ $mm$
radius-of-curvature mirror with a power reflectivity $99.8\%$ for
$1064$ $nm$ in the concave and a total transmissivity $70\%$ for
$532$ $nm$, which is mounted upon a PZT to adjust the cavity length.
The output wave is extracted from M2, which is a $30-mm$
radius-of-curvature mirror with a total transmissivity $3.3\%$ for
$1064$ $nm$ and a reflectivity $99\%$ for $532$ $nm$ in the concave.
Due to the large transmission of input coupler at $532$ $nm$, the
pump field can be thought as only passes the cavity twice without
resonation. The measured cavity finesse
was $148$ with the PPKTP crystal, which indicates the total cavity loss of $%
4.24$\%. Due to the high nonlinear coefficient of PPKTP, the measured
threshold power is only $35$ $mW$.

First, we fix the frequency of the subharmonic and the pump field with $%
\omega _p=2\omega $ and scan cavity length, which corresponds to the
condition of case 1. Figure 5 shows the experimental results: (a) without
the pump field, (b) $\varphi =\pi /2$ and $\beta ^{in}/\beta _{th}^{in}=0.33$%
, (c) $\varphi =\pi /2$ and $\beta ^{in}/\beta _{th}^{in}=0.71$, (d) $%
\varphi =\pi /2$ and $\beta ^{in}/\beta _{th}^{in}=0.9$, (e)
$\varphi =\pi /2-0.07$ and $\beta ^{in}/\beta _{th}^{in}=0.9$, (f)
$\varphi =\pi /2+0.07$ and $\beta ^{in}/\beta _{th}^{in}=0.9$. It
can be seen that the experimental curves are in good agreement with
the theoretical results shown in Fig.2, which are obtained with the
experimental parameters.

Then, we fix the cavity length and frequency of the pump field and
scan the frequency of the subharmonic field by the AOM, which
corresponds to the condition of case 2. The output including the
subharmonic and idle field is detected by a photodiode. There is a
beat-note signal in the photocurrent with frequency proportional to
the detuning. The very narrow dip and peak appeared in a broad
Lorentzian profile are observed experimentally as shown in Fig.6.
The insets in Fig.6 show the enlarged narrow dip and peak by
reducing the scanned range of frequency, which present the square
shape. Because the
measurement of transmission profile is dynamic processes, the shape of $%
\delta $ function for the narrow dip and peak in the theoretical
model becomes square shape in experiment. The width of the square
shape is $\sim$2KHz which is estimated from the voltage on VCO
(Voltage-Controlled Oscillator) of AOM.

$Conclusion.$ --- We reported the theoretical and experimental
results of coherence phenomena in the phase-sensitive optical
parametric amplification inside a cavity. The splitting in
transmission spectra of OPA was observed. Mode splitting, as well
known, occurs not only in coupled quantum system, but also in
coupled optical resonators and in coupled mechanical and electronic
oscillators. To the best of our knowledge, we first observed mode
splitting experimentally in the optical parametric process. This
system will be important for practically optical and photonic
applications such as optical filters, delay lines, and closely
relate to the coherent phenomenon of EIT predicted for quantum
systems. OPA also has a important application as squeezed light
source. Our results may help us to investigate quantum noise
spectrum.

$^{\dagger} $Corresponding author's email address: jzhang74@yahoo.com,
jzhang74@sxu.edu.cn

\smallskip \acknowledgments
J. Zhang thanks Prof. Kunchi Peng and Changde Xie for the helpful
discussions.This research was supported in part by National Natural
Science Foundation of China (Approval No.60178012), Program for New
Century Excellent Talents in University, Natural Science Foundation
of Shanxi Province, and the Research Fund for the Returned Abroad
Scholars of Shanxi Province.

REFERENCES

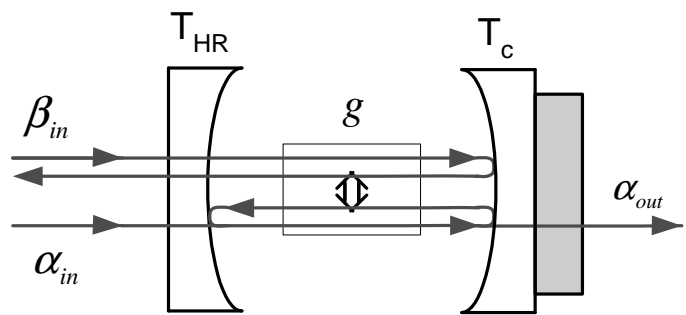
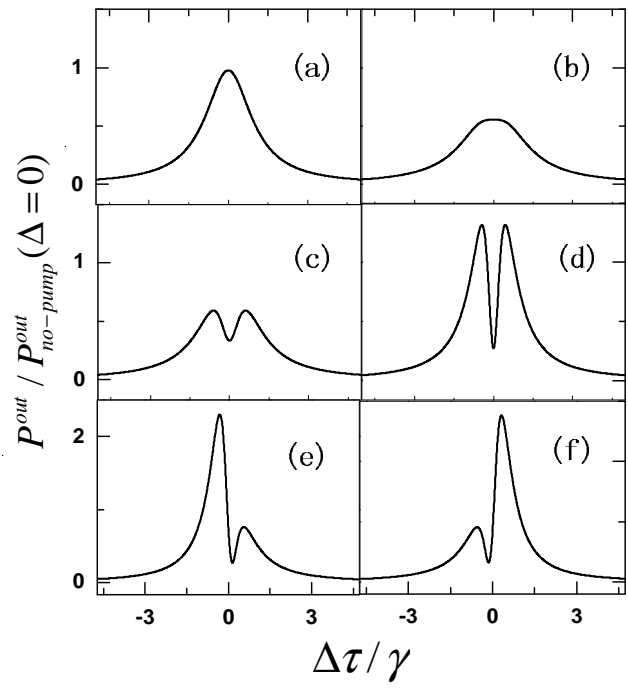
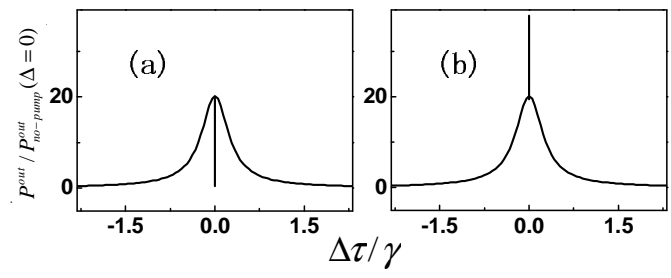
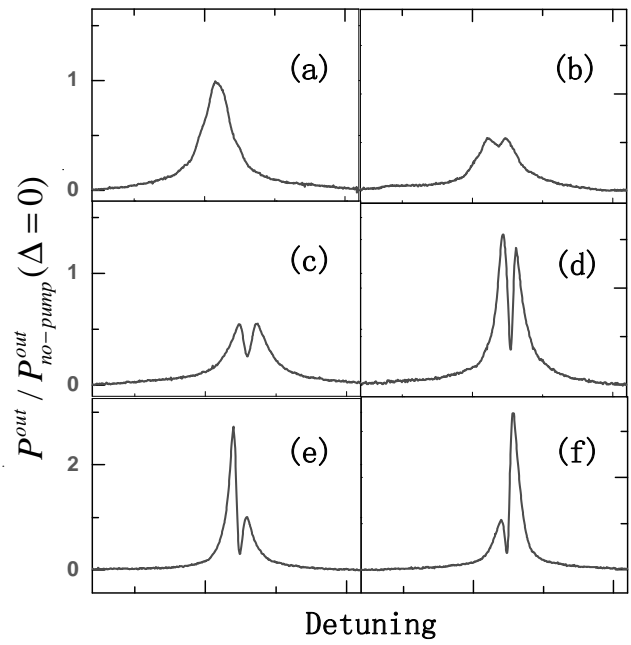
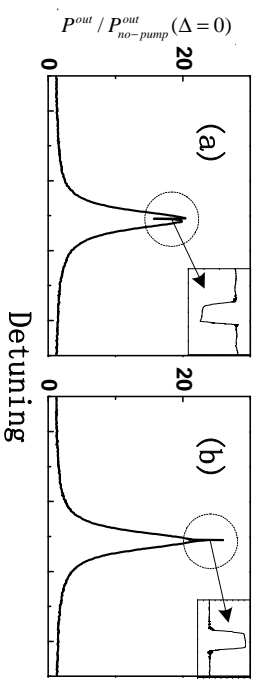
\end{document}